\documentclass[11pt,letterpaper,reqno]{amsart}

\title[Symmetric Representation of Rigid Body Equations and Symplectic Reduction]{The Symmetric Representation of the Generalized Rigid Body Equations and Symplectic Reduction}
\author{Tomoki Ohsawa}
\address{Department of Mathematical Sciences, The University of Texas at Dallas, 800 W Campbell Rd, Richardson, TX 75080-3021}
\email{tomoki@utdallas.edu}

\date{\today}



\usepackage{graphicx,amssymb,paralist}
\usepackage[margin=1in, marginpar=.5in]{geometry}

\usepackage[numbers,sort&compress]{natbib}
\usepackage[colorlinks=false]{hyperref}

\usepackage{tikz-cd}

\theoremstyle{plain}
\newtheorem{theorem}{Theorem}
\newtheorem*{theorem*}{Theorem}

\theoremstyle{definition}

\theoremstyle{remark}
\newtheorem{remark}[theorem]{Remark}

\def\od#1#2{\dfrac{d#1}{d#2}}

\def\parentheses#1{{\left(#1\right)}}

\def\tr{\mathop{\mathrm{tr}}\nolimits}

\def\rank{\mathop{\mathrm{rank}}\nolimits}

\def\R{\mathbb{R}}

\def\N{\mathbb{N}}
\def\defeq{\mathrel{\mathop:}=}
\def\eqdef{=\mathrel{\mathop:}}
\def\setdef#1#2{{\left\{ #1 \ |\ #2 \right\}}}
\def\ip#1#2{{\left\langle#1,#2\right\rangle}}

\def\diag{\operatorname{diag}}

\def\Mat{\mathsf{M}}

\def\SO{\mathsf{SO}}

\def\Sp{\mathsf{Sp}}

\def\Orth{\mathsf{O}}
\def\orth{\mathfrak{o}}

\def\so{\mathfrak{so}}
\def\sp{\mathfrak{sp}}

\def\Mat{\mathsf{M}}

\newenvironment{tbmatrix}{\left[\begin{smallmatrix}}{\end{smallmatrix}\right]}

\def\d{{\bf d}}
\def\ins#1{{\bf i}_{#1}}

\newcommand\Ad{\operatorname{Ad}}
\newcommand\ad{\operatorname{ad}}

\begin{document}

\footskip=.6in

\begin{abstract}
  We show that a symplectic reduction of the symmetric representation of the generalized $n$-dimensional rigid body equations yields the $n$-dimensional Euler equation. This result provides an alternative to the more elaborate relationship between these equations established by Bloch, Crouch, Marsden, and Ratiu. Specifically, we exploit the inherent $\mathsf{Sp}(2n,\mathbb{R})$-symmetry in the symmetric representation to present its relationship with the Euler equation via symplectic reduction facilitated by the dual pair recently developed by Skerritt and Vizman.
\end{abstract}

\maketitle

\section{Introduction}
\subsection{Euler's Equation for Rigid Body Dynamics}
The configuration space for the rotational dynamics of a rigid body in $\R^{3}$ about its center of mass is the set of all three-dimensional rotations
\begin{equation*}
  \SO(3) \defeq \setdef{ Q \in \Mat_{3}(\R) }{ Q^{T}Q = I,\, \det Q = 1},
\end{equation*}
where $\Mat_{n}(\R)$ is the vector space of real $n \times n$ matrices.
Hence one can describe the dynamics as a Hamiltonian system on the cotangent bundle (or the phase space) $T^{*}\SO(3)$.
However, the presence of the $\SO(3)$-symmetry helps us reduce the system to the dual $\so(3)^{*} \cong (\R^{3})^{*}$ of the Lie algebra $\so(3) \cong \R^{3}$ of $\SO(3)$; one may think of $\so(3)$ and $\so(3)^{*}$ as the space of all possible values of the body angular velocity and momentum, respectively, seen in the frame attached to the body.
This yields Euler's equation for the body angular momentum $\Pi$ in $\so(3)^{*}$ (identified with $\R^{3}$):
\begin{equation*}
  \dot{\Pi} = \Pi \times \mathbb{I}^{-1}\Pi,
\end{equation*}
where $\mathbb{I} \defeq \diag(I_{1}, I_{2}, I_{3})$ is the inertia tensor with respect to the principal axes of the body, and $\Omega \defeq \mathbb{I}^{-1}\Pi$ is the body angular velocity in $\so(3)$ (again identified with $\R^{3}$).

\subsection{Generalized Rigid Body Equations}
\label{ssec:Generalized_Rigid_Body}
\citet{Ma1976} and \citet{Ra1980} generalized the three-dimensional rigid body equations to $n$ dimensions as follows:
Let
\begin{equation*}
  \SO(n) \defeq \setdef{ Q \in \Mat_{n}(\R) }{ Q^{T}Q = I,\, \det Q = 1}
\end{equation*}
and $\so(n)$ be its Lie algebra, and equip $\so(n)$ with the inner product $\ip{\,\cdot\,}{\,\cdot\,}\colon \so(n) \times \so(n) \to \R$ defined as
\begin{equation}
  \label{eq:ip-so(n)}
  \ip{A}{B} \defeq \frac{1}{2}\tr(A^{T}B).
\end{equation}
So we may identify the dual $\so(n)^{*}$ with $\so(n)$.
Under this identification, we define the inertia operator $\mathcal{I}\colon \so(n) \to \so(n)^{*} \cong \so(n)$ of the generalized $n$-dimensional rigid body as
\begin{equation}
  \label{eq:mathcalI}
  \mathcal{I}(\Omega) \defeq \Lambda\Omega + \Omega\Lambda,
\end{equation}
where $\Lambda \defeq \diag(\lambda_{1}, \dots, \lambda_{n})$ with $\lambda_{i} + \lambda_{j} > 0$ for all $i \neq j$.
This is a generalization of the inertia tensor $\mathbb{I}$ defined above.
In fact, setting $I_{1} = \lambda_{2} + \lambda_{3}$, $I_{2} = \lambda_{3} + \lambda_{1}$, and $I_{3} = \lambda_{1} + \lambda_{2}$ for $n = 3$ yields
\begin{equation*}
 \Pi = \mathcal{I}(\Omega) =
  \begin{bmatrix}
    0 & -I_{3}\Omega_{3} & I_{2}\Omega_{2} \\
    I_{3}\Omega_{3} & 0 & -I_{1}\Omega_{1} \\
    -I_{2}\Omega_{2} & I_{1}\Omega_{1} & 0
  \end{bmatrix}
  \leftrightarrow
  \begin{bmatrix}
    I_{1}\Omega_{1} \\
    I_{2}\Omega_{2} \\
    I_{3}\Omega_{3}
  \end{bmatrix},
\end{equation*}
which is the body angular momentum in $\so(3)^{*} \cong \R^{3}$.
Let $Q \in \SO(n)$ be the rotational configuration of the generalized rigid body, $\Omega \defeq Q^{-1}\dot{Q} \in \so(n)$ be the body angular velocity, and $\Pi \defeq \mathcal{I}(\Omega) \in \so(n)^{*} \cong \so(n)$ be the body angular momentum.
Then the generalized rigid body or Euler--Poisson equations on $\SO(n) \times \so(n)^{*}$ are given by
\begin{subequations}
  \label{eq:RB}
  \begin{align}
    \dot{Q} &= Q \Omega, \\
    \dot{\Pi} &= [\Pi, \Omega], \label{eq:Euler}
  \end{align}
\end{subequations}
where $[\,\cdot\,,\,\cdot\,]\colon \so(n) \times \so(n) \to \so(n)$ is the standard commutator.
The $n$-dimensional Euler equation~\eqref{eq:Euler} is the Hamiltonian system on a coadjoint orbit in $\so(n)^{*}$ with the Hamiltonian (see \citet[Theorem~3.1]{Ra1980}):
\begin{equation}
  \label{eq:h}
  h(\Pi) \defeq \frac{1}{2}\ip{ \Pi }{ \mathcal{I}^{-1}(\Pi) }.
\end{equation}

\subsection{Symmetric Representation}
The so-called \textit{symmetric representation} of the generalized rigid body equations was originally discovered by \citet{BlCr1996} as a necessary condition of optimality for the following optimal control problem:
Let $T > 0$ be fixed, and consider
\begin{equation*}
  \min_{U} \int_{0}^{T} \frac{1}{2}\ip{\mathcal{I}(U)}{U}\,dt
  \quad \text{subject to}\quad
  \dot{Q} = Q U
  \text{ and }
  Q(0) \in \SO(n),
\end{equation*}
where $\mathcal{I}$ is the inertia operator~\eqref{eq:mathcalI} and $U \colon [0,T] \to \so(n)$.
Formally, the configuration space is $\Mat_{n}(\R)$, and thus $Q \colon [0,T] \to \Mat_{n}(\R)$.
However, we impose the initial condition $Q(0) \in \SO(n)$ so that $Q(t) \in \SO(n)$ for any $t \in [0,T]$ because of the constraint $Q(t)^{-1}\dot{Q}(t) = U(t) \in \so(n)$.
This is a special case of the so-called \textit{embedded optimal control problem} considered in \citet{BlCrNoSa2011}.

The control Hamiltonian $H_{\rm c}\colon T^{*}\Mat_{n}(\R) \times \so(n) \to \R$ is then defined as
\begin{equation*}
  H_{\rm c}(Q,P,U)
  \defeq P \cdot (Q U) - \frac{1}{2}\ip{\mathcal{I}(U)}{U} \\
  = \tr\!\parentheses{P^{T} Q U} - \frac{1}{2}\ip{\mathcal{I}(U)}{U},
\end{equation*}
where the ``$\,\cdot\,$'' above stands for the natural dual pairing
\begin{equation*}
  T^{*}_{Q}\Mat_{n}(\R) \times T_{Q}\Mat_{n}(\R) \to \R;
  \qquad
  (P, \dot{Q}) \mapsto \tr\parentheses{ P^{T}\dot{Q} } \eqdef P \cdot \dot{Q}.
\end{equation*}
By the Pontryagin Maximum Principle~\cite{PoBoGaMi1962}, the optimal control $U^{\star}$ necessarily maximizes the Hamiltonian, i.e., for any $\delta U \in \so(n)$,
\begin{equation*}
  \left.\od{}{s} H_{\rm c}(Q,P,U^{\star} + s\,\delta U) \right|_{s=0} = 0.
\end{equation*}
However,
\begin{align*}
  \left.\od{}{s} H_{\rm c}(Q,P,U^{\star} + s\,\delta U) \right|_{s=0}
  &= \tr\!\parentheses{P^{T} Q \delta U} - \ip{\mathcal{I}(U^{\star})}{\delta U} \\
  &= \frac{1}{2}\tr\!\parentheses{ \parentheses{ Q^{T}P - P^{T}Q }^{T} \delta U } - \ip{\mathcal{I}(U^{\star})}{\delta U} \\
  &= \ip{ Q^{T}P - P^{T}Q - \mathcal{I}(U^{\star}) }{ \delta U }.
  % &= \ip{ \frac{\delta H_{\rm c}}{\delta U}(Q,P,U^{\star}) }{ \delta U }
\end{align*}
Note that $Q^{T}P - P^{T}Q - \mathcal{I}(U^{\star}) \in \so(n)$.
Since $\delta U \in \so(n)$ is arbitrary, we have $Q^{T}P - P^{T}Q = \mathcal{I}(U^{\star})$, and thus obtain the optimal control $U^{\star}$ as follows:
\begin{equation*}
  U^{\star}(Q,P) = \mathcal{I}^{-1}\parentheses{ Q^{T}P - P^{T}Q }.
\end{equation*}
In what follows, we write $\Omega \defeq U^{\star}(Q,P)$ for short; this quantity in fact coincides with the angular velocity $\Omega$ defined in Section~\ref{ssec:Generalized_Rigid_Body} as we shall see below.
As a result, the (optimal) Hamiltonian $H\colon T^{*}\Mat_{n}(\R) \to \R$ is given by
\begin{equation}
  \label{eq:H-QP}
  H(Q,P) \defeq H_{\rm c}(Q,P,U^{\star}(Q,P)) = \frac{1}{2}\ip{ Q^{T}P - P^{T}Q }{ \mathcal{I}^{-1}\parentheses{ Q^{T}P - P^{T}Q } },
\end{equation}
whereas the standard symplectic form $\omega$ on $T^{*}\Mat_{n}(\R)$ is written as
\begin{equation}
  \label{eq:Omega-QP}
  \omega\parentheses{ (\dot{Q}_{1},\dot{P}_{1}), (\dot{Q}_{2},\dot{P}_{2}) }
  = \tr\!\parentheses{ \dot{Q}_{1}^{T} \dot{P}_{2} - \dot{P}_{1}^{T} \dot{Q}_{2} }.
\end{equation}
The optimal solution is necessarily an integral curve of the Hamiltonian vector field $X_{H}$ on $T^{*}\Mat_{n}(\R)$ defined by Hamilton's equation $\ins{X_{H}}\omega = \d{H}$, or in coordinates,
\begin{equation}
  \label{eq:SymRB}
  \dot{Q} = Q \Omega,
  \qquad
  \dot{P} = P \Omega.
\end{equation}
These equations are called the \textit{symmetric representation} of the generalized rigid body equations~\eqref{eq:RB}.
In fact, if we set
\begin{equation*}
  \Pi(t) \defeq \mathcal{I}(\Omega(t)) = Q(t)^{T} P(t) - P(t)^{T} Q(t)
\end{equation*}
for $t \in [0,T]$, then $\Pi\colon [0,T] \to \so(n)^{*} \cong \so(n)$ satisfies the $n$-dimensional Euler equation~\eqref{eq:Euler}; see also \citet{BlCrMaSa2008,BlCrNoSa2011} for its generalization to other matrix Lie groups.

Although it is a straightforward calculation to show that the generalized rigid body equations~\eqref{eq:RB} follow from the symmetric representation~\eqref{eq:SymRB}, the question remains as to how these two Hamiltonian systems are related to each other from the symplectic-geometric point of view.
\citet{BlCrMaRa2002} gave one such relationship: Specifically, they constructed symplectic submanifolds $S \subset \SO(n) \times \SO(n)$ of $T^{*}\Mat_{n}(\R) \cong \Mat_{n}(\R) \times \Mat_{n}(\R)$ and $S_{M}$ of $T^{*}\SO(n) \cong \SO(n) \times \so(n)^{*}$ as well as a diffeomorphism between $S$ and $S_{M}$ in a rather elaborate manner to establish an equivalence between the symmetric representation~\eqref{eq:SymRB} and the Euler--Poisson equations~\eqref{eq:RB}.

\subsection{Main Result}
We present an alternative connection between the symmetric representation~\eqref{eq:SymRB} and the Euler equation~\eqref{eq:Euler} by showing that the two are related via symplectic reduction.
Although this is not an equivalence, this connection exploits an inherent symmetry of the symmetric representation that seems to have been overlooked, and provides a geometrically natural and appealing alternative to the result of \citet{BlCrMaRa2002}.
More specifically, we show the following:
\begin{theorem*}
  The symmetric representation~\eqref{eq:SymRB} of the $n$-dimensional generalized rigid body equation possesses an $\Sp(2n,\R)$-symmetry, and the Marsden--Weinstein reduction~\cite{MaWe1974} of \eqref{eq:SymRB} restricted to an open subset of $T^{*}\Mat_{n}(\R)$ by the symmetry at a certain level set of the associated momentum map yields the $n$-dimensional Euler equation~\eqref{eq:Euler} in a coadjoint orbit in $\so(n)^{*}$.
\end{theorem*}

\section{Symmetry and Conservation Law in the Symmetric Representation}
\subsection{$\Sp(2n,\R)$-Symmetry of the Symmetric Representation}
Let us first identify the cotangent bundle $T^{*}\Mat_{n}(\R)$ with the vector space of real $2n \times n$ matrices:
\begin{equation*}
  T^{*}\Mat_{n}(\R)
  \cong \Mat_{2n\times n}(\R)
  = \setdef{
    Z = 
    \begin{bmatrix}
      Q \\
      P
    \end{bmatrix}
  }{ Q, P \in \Mat_{n}(\R) }.
\end{equation*}
Then we may write the symplectic form~\eqref{eq:Omega-QP} in a more succinct form:
For any $Z \in \Mat_{2n\times n}(\R)$ and any $X, Y \in T_{Z}\Mat_{2n\times n}(\R)$,
\begin{equation}
  \label{eq:Omega}
  \omega(X, Y)
  = \tr\!\parentheses{ X^{T} \mathbb{J} Y }
  \quad\text{with}\quad
  \mathbb{J} \defeq
  \begin{bmatrix}
    0 & I \\
    -I & 0
  \end{bmatrix}
\end{equation}
where $I$ is the $n \times n$ identity matrix.
Notice also that we may rewrite the Hamiltonian $H\colon \Mat_{2n\times n}(\R) \to \R$ defined in \eqref{eq:H-QP} in the following more concise form:
\begin{equation}
  \label{eq:H}
  H(Z) = \frac{1}{2}\ip{ Z^{T}\mathbb{J}Z }{ \mathcal{I}^{-1}(Z^{T}\mathbb{J}Z) }.
\end{equation}

Let $\Sp(2n,\R)$ be the symplectic group
\begin{equation*}
  \Sp(2n,\R) \defeq
  \setdef{
    S \in \Mat_{2n}(\R)
  }{ S^{T} \mathbb{J} S = \mathbb{J}},
\end{equation*}
and consider the $\Sp(2n,\R)$-action on $\Mat_{2n\times n}(\R)$ by left multiplication, i.e.,
\begin{equation}
  \label{eq:Phi}
  \Phi\colon \Sp(2n,\R) \times \Mat_{2n\times n}(\R) \to \Mat_{2n\times n}(\R);
  \qquad
  (S,Z) \mapsto S Z \eqdef \Phi_{S}(Z).
\end{equation}
Then one easily sees that this action is symplectic as well as that the Hamiltonian is invariant under the action, i.e., $\Phi_{S}^{*}\omega = \omega$ and $H \circ \Phi_{S} = H$ for any $S \in \Sp(2n,\R)$.

\subsection{$\Sp(2n,\R)$-Momentum Map}
Let us find the momentum map associated with the above action $\Phi$.
First notice that the symplectic form~\eqref{eq:Omega} is written as $\omega = -\d\theta$ with the one-form $\theta$ on $\Mat_{2n\times n}(\R)$ defined as follows:
For any $Z =
\begin{tbmatrix}
  Q \smallskip\\
  P
\end{tbmatrix}
\in \Mat_{2n\times n}(\R)$ and any $\dot{Z} = \begin{tbmatrix}
  \dot{Q} \smallskip\\
  \dot{P}
\end{tbmatrix}\in T_{Z}\Mat_{2n\times n}(\R)$,
\begin{equation*}
  \theta(Z)\cdot\dot{Z} \defeq -\frac{1}{2}\tr\!\parentheses{ Z^{T}\mathbb{J}\dot{Z} }
  = \frac{1}{2}\parentheses{ P^{T}\dot{Q} - Q^{T}\dot{P} }.
\end{equation*}
It is clear that $\theta$ is invariant under the $\Sp(2n,\R)$-action, i.e., $\Phi_{S}^{*}\theta = \theta$ for any $S \in \Sp(2n,\R)$.

Let $\sp(2n,\R)$ be the Lie algebra of $\Sp(2n,\R)$, i.e.,
\begin{equation*}
  \sp(2n,\R) = \setdef{ \xi \in \Mat_{2n}(\R) }{ \xi^{T}\mathbb{J} + \mathbb{J}\xi = 0 }.
\end{equation*}
We equip $\sp(2n,\R)$ with the inner product $\ip{\,\cdot\,}{\,\cdot\,}\colon \sp(2n,\R) \times \sp(2n,\R) \to \R$ defined as
\begin{equation*}
  \ip{\xi}{\eta} \defeq \frac{1}{2}\tr(\xi^{T} \eta)
\end{equation*}
just as in \eqref{eq:ip-so(n)}, and thus we may identify the dual $\sp(2n,\R)^{*}$ with $\sp(2n,\R)$.
The infinitesimal generator of the above action $\Phi$ is, for any $\xi \in \sp(2n,\R)$,
\begin{equation*}
  \xi_{\Mat_{2n\times n}(\R)}(Z) \defeq \left.\od{}{s} \Phi_{\exp(s\xi)}(Z) \right|_{s=0} = \xi Z.
\end{equation*}

Since $T^{*}\Mat_{n}(\R) \cong \Mat_{2n\times n}(\R)$ is an exact symplectic manifold with $\omega = -\d\theta$ and $\Phi$ leaves $\theta$ invariant, the associated momentum map $\mathbf{J}\colon \Mat_{2n\times n}(\R) \to \sp(2n,\R)^{*} \cong \sp(2n,\R)$ satisfies the following (see, e.g., \citet[Theorem~4.2.10 on p.~282]{AbMa1978}):
For any $\xi \in \sp(2n,\R)$,
\begin{align*}
  \ip{ \mathbf{J}(Z) }{ \xi }
  &= \theta(Z) \cdot \xi_{\Mat_{2n\times n}(\R)}(Z) \\
  &= -\frac{1}{2}\tr\!\parentheses{ Z^{T}\mathbb{J} \xi Z } \\
%  &= -\frac{1}{2}\tr\!\parentheses{ Z Z^{T}\mathbb{J} \xi } \\
  &= \frac{1}{2}\tr\!\parentheses{ (\mathbb{J} Z Z^{T})^{T} \xi } \\
  &= \ip{ \mathbb{J} Z Z^{T} }{ \xi },
\end{align*}
and so we obtain
\begin{equation}
  \label{eq:J}
  \mathbf{J}(Z) = \mathbb{J} Z Z^{T} =
  \begin{bmatrix}
    P Q^{T} & P P^{T} \\
    -Q Q^{T} & -Q P^{T}
  \end{bmatrix}.
\end{equation}
This is the special case with $m = n$ of \citet[Proposition~4.1]{SkVi2019}.
It is also easy to see that $\mathbf{J}$ is equivariant:
For any $S \in \Sp(2n,\R)$,
\begin{equation*}
  \mathbf{J} \circ \Phi_{S} = \Ad_{S^{-1}}^{*} \circ \mathbf{J}.
\end{equation*}

By Noether's Theorem (see, e.g., \citet[Theorem~11.4.1 on p.~372]{MaRa1999}), $\mathbf{J}$ is a conserved quantity of the symmetric representation~\eqref{eq:SymRB} due to the $\Sp(2n,\R)$-symmetry.
That each block of this matrix is a conserved quantity is also pointed out by \citet{BlCrMaRa2002} via direct computations.

\section{Symplectic Reduction of the Symmetric Representation}
\subsection{Symplectic Reduction}
Let $P$ be a symplectic manifold with symplectic form $\omega$, and suppose in addition that there is a symplectic action of a Lie group $\mathsf{G}$ on $P$, $\mathfrak{g}^{*}$ be the dual of the Lie algebra $\mathfrak{g}$ of $\mathsf{G}$, and $\mathbf{J}\colon P \to \mathfrak{g}^{*}$ be the momentum map associated with the action.
The Marsden--Weinstein reduction~\cite{MaWe1974} (see also \cite[Sections~1.1 \& 1.2]{MaMiOrPeRa2007}) states that, if either (i)~the $\mathsf{G}$-action on $P$ is free and proper, or (ii)~$\mu \in \mathfrak{g}^{*}$ is a regular value of $\mathbf{J}$ and the action of the isotropy group
\begin{equation*}
  \mathsf{G}_{\mu} \defeq \setdef{
    g \in \mathsf{G}
  }{
    \Ad_{g^{-1}}^{*} \mu = \mu
  }
\end{equation*}
on the level set $\mathbf{J}^{-1}(\mu)$ is free and proper, then the quotient space $\mathbf{J}^{-1}(\mu)/\mathsf{G}_{\mu}$ is also a symplectic manifold with symplectic structure $\overline{\omega}_{\mu}$ that is naturally induced by $\omega$ and the geometric setting; see below for more details.
Now, given a Hamiltonian $H\colon P \to \R$, one may define the Hamiltonian vector field $X_{H}$ on $P$ by setting $\ins{X_{H}}\omega = \d{H}$.
If $H$ is invariant under the $\mathsf{G}$-action, it gives rise to the reduced Hamiltonian $\overline{H}_{\mu}$ on $\mathbf{J}^{-1}(\mu)/\mathsf{G}_{\mu}$, and then the Hamiltonian dynamics in $P$ is reduced to the Hamiltonian dynamics in $\mathbf{J}^{-1}(\mu)/\mathsf{G}_{\mu}$ defined in terms of $\overline{H}_{\mu}$ and $\overline{\omega}_{\mu}$.
In other words, one can reduce a Hamiltonian system with a Lie-group symmetry to a lower-dimensional Hamiltonian system.

\subsection{Some Technical Issues of Symplectic Reduction}
\label{ssec:technical_issues}
We would like to perform the Marsden--Weinstein reduction of the symmetric representation~\eqref{eq:SymRB}; here we have $P = T^{*}\Mat_{n}(\R) \cong \Mat_{2n\times n}(\R)$, $\mathsf{G} = \Sp(2n,\R)$ with the action $\Phi$ defined in \eqref{eq:Phi}, and the momentum map $\mathbf{J}$ from \eqref{eq:J}.
However, condition~(i) clearly does not hold because $\Phi$ is not a free action on $\Mat_{2n\times n}(\R)$, and so one either needs to remedy this or check (ii).
Otherwise, the quotient $\mathbf{J}^{-1}(\mu)/\Sp(2n,\R)_{\mu}$ may not be a manifold.
The other issue is how to characterize the quotient space $\mathbf{J}^{-1}(\mu)/\Sp(2n,\R)_{\mu}$ explicitly in order to describe the symplectic structure and the reduced dynamics there in an explicit manner.

Fortunately, the recent work by \citet{SkVi2019} provides a geometric setting that is tailor-made for circumventing these issues.
More specifically, we consider the following pair of momentum maps defined on $\Mat_{2n\times n}(\R)$:
\begin{equation*}
  \begin{tikzcd}
    \sp(2n,\R)^{*} & \Mat_{2n\times n}(\R) \arrow[swap]{l}{\mathbf{J}} \arrow{r}{\mathbf{M}} & \orth(n)^{*},
  \end{tikzcd}
\end{equation*}
where $\mathbf{J}$ is defined above in \eqref{eq:J} and $\mathbf{M}$ is the momentum map associated with the action of the orthogonal group $\Orth(n)$ on $\Mat_{2n\times n}(\R)$ to be described below; $\orth(n)$ is the Lie algebra of $\Orth(n)$.
What they show is that, by considering an open subset $\mathcal{Z}$ of $\Mat_{2n\times n}(\R)$ and restricting the actions and the momentum maps there, one may identify the Marsden--Weinstein quotient $\mathbf{J}^{-1}(\mu)/\Sp(2n,\R)_{\mu}$ with a coadjoint orbit in $\orth(n)^{*}$.
We note that their result is slightly more general than this: They have the result with $\Mat_{2n\times m}(\R)$ and $\Orth(m)$ with $n, m \in \N$ instead, and so our setting is the special case of theirs with $m = n$.

\subsection{$\Orth(n)$-action and Momentum Map}
Consider the action of the orthogonal group $\Orth(n)$ on $\Mat_{2n\times n}(\R)$ defined by right multiplication, i.e.,
\begin{equation*}
  %\label{eq:Psi}
  \Psi\colon \Orth(n) \times \Mat_{2n\times n}(\R) \to \Mat_{2n\times n}(\R);
  \qquad
  (R, Z) \mapsto Z R = \Psi_{R}(Z).
\end{equation*}
It is a straightforward calculation to see that $\Psi$ leaves the one-form $\theta$ invariant and hence is a symplectic action with respect to the symplectic form $\omega$ defined in \eqref{eq:Omega}, i.e., $\Psi_{R}^{*}\theta = \theta$ and hence $\Psi_{R}^{*}\omega = \omega$ for any $R \in \Orth(n)$.

Since $\orth(n) = \so(n)$, we identify the dual $\orth(n)^{*}$ with $\orth(n)$ via the inner product~\eqref{eq:ip-so(n)}.
Then, following a similar calculation as the one for $\mathbf{J}$ from above (see also \citet[Proposition~4.1]{SkVi2019}), we obtain the associated momentum map $\mathbf{M} \colon \Mat_{2n\times n}(\R) \to \orth(n)^{*} \cong \orth(n)$ as follows:
\begin{equation}
  \label{eq:M}
  \mathbf{M}(Z) = Z^{T} \mathbb{J} Z = Q^{T}P - P^{T}Q.
\end{equation}
Again, it is a straightforward calculation to see that $\mathbf{M}$ is equivariant, i.e., for any $R \in \Orth(n)$,
\begin{equation*}
  \mathbf{M} \circ \Psi_{R} = \Ad_{R}^{*} \circ \mathbf{M}.
\end{equation*}

\subsection{Symplectic Reduction and Dual Pair}
%\label{ssec:dual_pair}
Following \citet{SkVi2019}, let us consider the subset of $\Mat_{2n\times n}(\R)$ that is consisting of the full-rank elements, i.e.,
\begin{equation}
  \label{eq:mathcalZ}
  \mathcal{Z} \defeq \setdef{ Z \in \Mat_{2n\times n}(\R) }{ \rank Z = n }.
\end{equation}
As shown in \cite{SkVi2019}, $\mathcal{Z}$ is an open subset of $\Mat_{2n\times n}(\R)$, and the actions $\Phi$ and $\Psi$ preserve $\mathcal{Z}$.
Hence we may restrict the symplectic form $\omega$ and the momentum maps $\mathbf{J}$ and $\mathbf{M}$ to $\mathcal{Z}$; we denote these restrictions by the same symbols for simplicity of notation:
\begin{equation*}
  %\label{eq:dual_pair}
  \begin{tikzcd}
    \sp(2n,\R)^{*} & \mathcal{Z} \arrow[swap]{l}{\mathbf{J}} \arrow{r}{\mathbf{M}} & \orth(n)^{*}.
  \end{tikzcd}
\end{equation*}

\citet[Proposition~4.2]{SkVi2019} proved that $\Phi$ and $\Psi$ define \textit{mutually transitive actions} on $\mathcal{Z}$ in the following sense:
(i)~The $\Sp(2n,\R)$-action~$\Phi$ and the $\Orth(n)$-action $\Psi$ commute; (ii)~$\Phi$ and $\Psi$ are symplectic actions; (iii)~the momentum maps $\mathbf{J}$ and $\mathbf{M}$ are equivariant; (iv)~each level set of $\mathbf{J}$ is an $\Orth(n)$-orbit, and each level set of $\mathbf{M}$ is an $\Sp(2n,\R)$-orbit.

The mutual transitivity has the following important consequence~(see also \citet[Theorem 2.9~(iii)]{BaWu2012} and \citet[Proposition~3.5]{Sk2018}):
Let $Z_{0} \in \mathcal{Z}$ and $\mu_{0} \defeq \mathbf{J}(Z_{0})$ and $\Pi_{0} \defeq \mathbf{M}(Z_{0})$.
Then one can identify the Marsden--Weinstein quotient
\begin{equation*}
  \overline{\mathcal{Z}}_{\mu_{0}} \defeq \mathbf{J}^{-1}(\mu_{0})/\Sp(2n,\R)_{\mu_{0}}  
\end{equation*}
with the coadjoint orbit $\mathcal{O}_{\Pi_{0}}$ through $\Pi_{0}$ in $\orth(n)^{*}$.
\begin{remark}
  \label{rem:reduced_space}
  We do not have to check that the condition (mentioned in Section~\ref{ssec:technical_issues}) that the $\Sp(n,\R)$-action $\Phi$ or the $\Sp(n,\R)_{\mu_{0}}$ action on $\mathbf{J}^{-1}(\mu_{0})$ is free and proper.
  In fact, the smooth structure on the reduced space $\overline{\mathcal{Z}}_{\mu_{0}}$ is induced by that of the coadjoint orbit $\mathcal{O}_{\Pi_{0}}$.
  See the proof of Proposition~2.8 in \cite{SkVi2019} for details.
\end{remark}
More specifically, let $i_{\mu_{0}}\colon \mathbf{J}^{-1}(\mu_{0}) \hookrightarrow \mathcal{Z}$ be the inclusion and $\pi_{\mu_{0}}\colon \mathbf{J}^{-1}(\mu_{0}) \to \overline{\mathcal{Z}}_{\mu_{0}}$ be the quotient map.
Then the reduced symplectic form $\overline{\omega}_{\mu_{0}}$ on $\overline{\mathcal{Z}}_{\mu_{0}}$ is uniquely determined by
\begin{equation*}
  i_{\mu_{0}}^{*} \omega = \pi_{\mu_{0}}^{*} \overline{\omega}_{\mu_{0}};
\end{equation*}
see \cite{MaWe1974} and \citet[Sections~1.1 \& 1.2]{MaMiOrPeRa2007}.
Also, let $\omega_{\mathcal{O}_{\Pi_{0}}}$ be the $(-)$-Kirillov--Kostant--Souriau symplectic structure, i.e., for any $\Pi \in \mathcal{O}_{\Pi_{0}}$ and $A, B \in \orth(n)$,
\begin{equation*}
  %\label{eq:KKS}
  \omega_{\mathcal{O}_{\Pi_{0}}}( -\ad_{A}^{*}\Pi, -\ad_{B}^{*}\Pi ) \defeq -\ip{\Pi}{[A,B]},
\end{equation*}
where $[\,\cdot\,,\,\cdot\,]$ is the commutator on $\orth(n)$; see, e.g., \citet[Chapter~1]{Ki2004} and \citet[Chapter~14]{MaRa1999}.
Then the momentum map $\mathbf{M}$ restricted to the level set $\mathbf{J}^{-1}(\mu_{0})$ gives rise to a diffeomorphism $\overline{\mathbf{M}}\colon \overline{\mathcal{Z}}_{\mu_{0}} \to \mathcal{O}_{\Pi_{0}}$; moreover this map is symplectic with respect to the above symplectic forms, i.e.,
\begin{equation*}
  \overline{\mathbf{M}}^{*}  \omega_{\mathcal{O}_{\Pi_{0}}} = \overline{\omega}_{\mu_{0}}.
\end{equation*}
The diagram below gives an overview of this result.
\begin{equation*}
  \begin{tikzcd}[column sep=7ex, row sep=7ex]
    \mathcal{Z} & \\
    \mathbf{J}^{-1}(\mu_{0}) \arrow{u}{i_{\mu_{0}}} \arrow[swap]{d}{\pi_{\mu_{0}}} \arrow{dr}{\mathbf{M}|_{\mathbf{J}^{-1}(\mu_{0})}} & \\
    \overline{\mathcal{Z}}_{\mu_{0}} \arrow[swap]{r}{\overline{\mathbf{M}}} & \mathcal{O}_{\Pi_{0}}
  \end{tikzcd}
\end{equation*}

\subsection{Reduction of Symmetric Representation}
Let $Q(0) = Q_{0} \in \SO(n)$ and $\Pi_{0} \in \orth(n)^{*}$ be the initial rotational configuration and the initial body angular momentum of the rigid body, and fix $P_{0} \in \SO(n)$ so that
\begin{equation*}
  Q_{0}^{T} P_{0} - P_{0}^{T} Q_{0} = \Pi_{0}.
\end{equation*}
See \citet{BlCr1996} and \citet{BlCrMaRa2002} for the condition under which this is possible.
Then clearly $Z_{0} \defeq (Q_{0}, P_{0})$ is in the open subset $\mathcal{Z} \subset \Mat_{2n\times n}(\R)$ defined in \eqref{eq:mathcalZ} and $\Pi_{0} = \mathbf{M}(Z_{0})$.
Now, setting
\begin{equation*}
  \mu_{0} \defeq \mathbf{J}(Z_{0}) = 
  \begin{bmatrix}
    P_{0} Q_{0}^{T} & I \smallskip\\
    -I & -Q_{0} P_{0}^{T}
  \end{bmatrix} \in \sp(2n,\R)^{*},
\end{equation*}
the level set
\begin{equation*}
  \mathbf{J}^{-1}(\mu_{0})
  = \setdef{
    \begin{bmatrix}
      Q \\
      P
    \end{bmatrix}
    \in \mathcal{Z}
  }{
    Q Q^{T} = I,\,
    P P^{T} = I,\,
    P Q^{T} = P_{0} Q_{0}^{T}
  }
\end{equation*}
is an invariant submanifold of the symmetric representation~\eqref{eq:SymRB}.

Let $h\colon \orth(n)^{*} \to \R$ be a collective Hamiltonian, i.e., $h \circ \mathbf{M} = H$.
From the expressions \eqref{eq:H} and \eqref{eq:M} of $H$ and $\mathbf{M}$, we find
\begin{equation*}
  h(\Pi) = \frac{1}{2}\ip{ \Pi }{ \mathcal{I}^{-1}(\Pi) },
\end{equation*}
which is the Hamiltonian~\eqref{eq:h} of the generalized rigid body in the body representation.
Then the result from the previous subsection implies that the $\Sp(2n,\R)$-reduced dynamics in $\overline{\mathcal{Z}}_{\mu_{0}}$ is equivalent to the Lie--Poisson equation
\begin{equation*}
  \dot{\Pi} = \ad_{Dh(\Pi)}^{*} \Pi
\end{equation*}
in the coadjoint orbit $\mathcal{O}_{\Pi_{0}} \subset \orth(n)^{*}$, where $Dh(\Pi) \in \orth(n)$ is defined so that, for any $\delta\Pi \in \orth(n)^{*}$,
\begin{equation*}
  \ip{ \delta\Pi }{ Dh(\Pi) } = \left.\od{}{s} h(\Pi + s\,\delta\Pi) \right|_{s=0}
  = \ip{ \delta\Pi }{ \mathcal{I}^{-1}(\Pi) },
\end{equation*}
that is, $Dh(\Pi) = \mathcal{I}^{-1}(\Pi)$.
However, under the identification $\orth(n)^{*} \cong \orth(n)$, $\ad_{A}^{*}\Pi = [\Pi, A]$ for any $A \in \orth(n)$ and $\Pi \in \orth(n)^{*}$, and thus we obtain
\begin{equation*}
  \dot{\Pi} = \left[ \Pi, \mathcal{I}^{-1}(\Pi) \right],
\end{equation*}
which is the $n$-dimensional Euler equation~\eqref{eq:Euler}.

\section*{Acknowledgments}
I would like to thank Paul Skerritt for helpful discussions on dual pairs.
This work was partially supported by NSF grant CMMI-1824798.

\bibliography{SymRepRigidBody}
\bibliographystyle{plainnat}

\end{document}